\documentclass[preprint2]{emulateapj}
\usepackage{cancel}	
\usepackage{color}

\def\M{M$_{\odot}$}

\def\og{OGLE13-079}

\shorttitle{OGLE-2013-SN-079}
\shortauthors{C. Inserra}

\begin{document}


\title{OGLE-2013-SN-079: a lonely supernova consistent with a helium shell detonation}


\author{C. Inserra\altaffilmark{1},
S. A. Sim\altaffilmark{1},
L. Wyrzykowski\altaffilmark{2,3},
S. J. Smartt\altaffilmark{1},
M. Fraser\altaffilmark{3},
M. Nicholl\altaffilmark{1},
K. J. Shen\altaffilmark{4},
A. Jerkstrand\altaffilmark{1},
 A. Gal-Yam\altaffilmark{5}, 
 D. A. Howell\altaffilmark{6,7}, 
K. Maguire\altaffilmark{8},
P. Mazzali\altaffilmark{9,10,11}, 
 S. Valenti\altaffilmark{6,7},   
S. Taubenberger\altaffilmark{11},
S. Benitez-Herrera\altaffilmark{11},
D. Bersier\altaffilmark{9}, 
N. Blagorodnova\altaffilmark{3},
H. Campbell\altaffilmark{3}, 
T.-W. Chen\altaffilmark{1}, 
N. Elias-Rosa\altaffilmark{10},
W. Hillebrandt\altaffilmark{11},
Z. Kostrzewa-Rutkowska\altaffilmark{2}, 
S. Koz\l owski\altaffilmark{2}, 
M. Kromer\altaffilmark{12},
J. D. Lyman\altaffilmark{13},
J. Polshaw\altaffilmark{1}, 
F. K. R\"opke\altaffilmark{14},
A. J. Ruiter\altaffilmark{15},
 K. Smith\altaffilmark{1}, S. Spiro\altaffilmark{16}, M. Sullivan\altaffilmark{17}, O. Yaron\altaffilmark{5}, D. Young\altaffilmark{1}
and F. Yuan\altaffilmark{15}
}
\altaffiltext{1}{Astrophysics Research Centre, School of Mathematics and Physics, Queens University
  Belfast, Belfast BT7 1NN, UK; c.inserra@qub.ac.uk}
\altaffiltext{2}{University of Warsaw, Astronomical Observatory, Al. Ujazdowskie 400-478 Warszawa, Poland}
\altaffiltext{3}{Institute of Astronomy, University of Cambridge, Madingley Road, CB3 0HA Cambridge, UK}
\altaffiltext{4}{Department of Astronomy and Theoretical Astrophysics Center, University of California, Berkeley, CA 94720, USA}
\altaffiltext{5}{Benoziyo Center for Astrophysics, Weizmann Institute of Science, 76100 Rehovot, Israel}
\altaffiltext{6}{Las Cumbres Observatory Global Telescope Network, 6740 Cortona Dr., Suite 102 Goleta, Ca 93117}
\altaffiltext{7}{Department of Physics, University of California, Santa Barbara, Broida Hall, Mail Code 9530, Santa Barbara, CA 93106-9530, USA}
\altaffiltext{8}{European Southern Observatory for Astronomical Research in the Southern Hemisphere (ESO), Karl-Schwarzschild-Str. 2, 85748 Garching b. Munchen, Germany}
\altaffiltext{9}{Astrophysics Research Institute, Liverpool John Moores University ,Liverpool, UK}
\altaffiltext{10}{INAF - Osservatorio Astronomico di Padova, Vicolo dell'Osservatorio 5, I-35122 Padova, Italy}
\altaffiltext{11}{Max-Planck-Institut f\"ur Astrophysik, Karl-Schwarzschild-Str. 1, 85741 Garching, Germany}
\altaffiltext{12}{The Oskar Klein Centre, Stockholm University, AlbaNova, SE-106 91 Stockholm, Sweden}
\altaffiltext{13}{Department of Physics, University of Warwick, Coventry CV4 7AL, UK}
\altaffiltext{14}{Institut f\"ur Theoretische Physik und Astrophysik, Universit\"at W\"urzburg  Emil-Fischer-Stra{\ss}e 31, D-97074 W\"urzburg, Germany}
\altaffiltext{15}{Research School of Astronomy \& Astrophysics, Mount Stromlo Observatory, The Australian National University
Cotter Rd., Weston Creek, ACT 2611, Australia}
\altaffiltext{16}{Department of Physics (Astrophysics), University of Oxford, DWB, Keble Road, Oxford OX1 3RH, UK}
\altaffiltext{17}{School of Physics and Astronomy, University of Southampton, Southampton, SO17 1BJ, UK}






\begin{abstract}

We present observational data for a peculiar supernova discovered by the OGLE-IV survey and followed by the Public ESO Spectroscopic Survey for Transient Objects. The inferred redshift of $z=0.07$ implies an absolute magnitude in the rest-frame $I$-band of M$_{I}\sim-17.6$ mag. This places it in the luminosity range between normal Type Ia SNe and novae. Optical and near infrared spectroscopy reveal mostly Ti and Ca lines, and an unusually red color arising from strong depression of flux at rest wavelengths $<5000$ \AA\/. To date, this is the only reported SN showing Ti-dominated spectra. 
The data are broadly consistent with existing models for the pure detonation of a helium shell around a low-mass CO white dwarf and ``double-detonation" models that include a secondary detonation of a CO core following a primary detonation in an overlying helium shell. 

 \end{abstract}


\section{Introduction}\label{sec:intro}

\begin{figure*}
\center
\includegraphics[width=18cm]{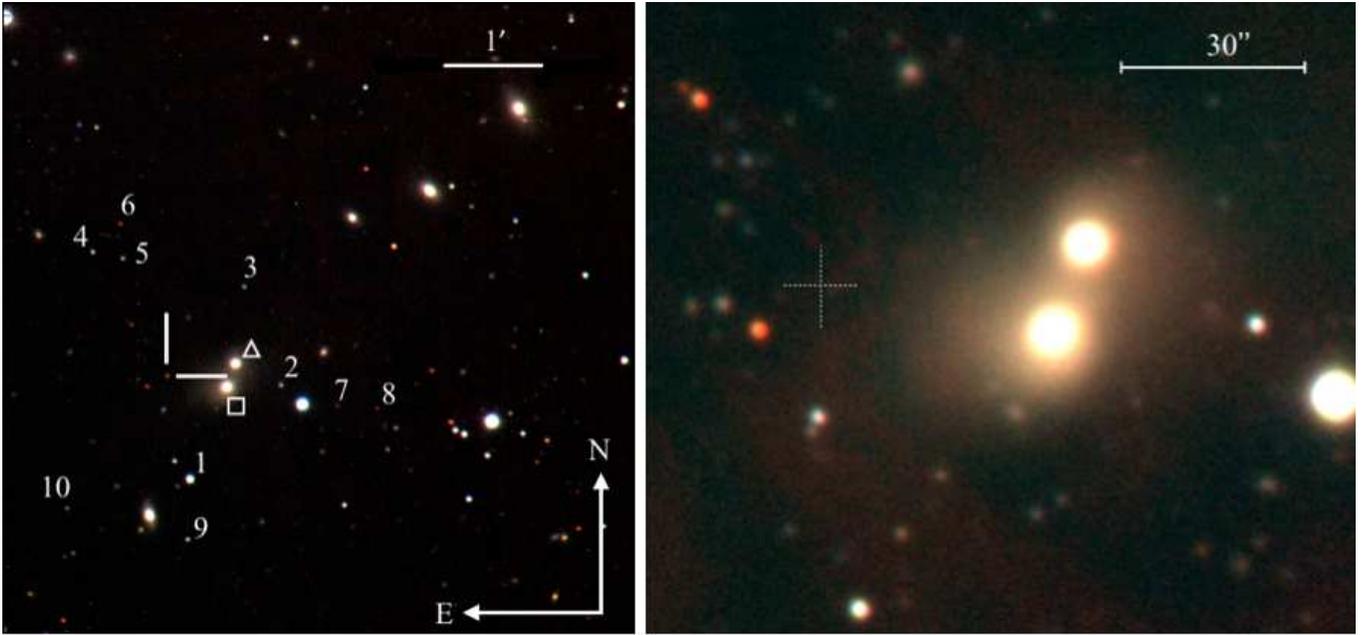}
\caption{Left: FTS+Merope+$g/r/i$ 
image of \og\/ (cross marks). 
The sequence of stars in the field is indicated.
The two closest galaxies to  \og\/  are indicated (square and triangle). 
Right:  NTT+EFOSC2+$g/r/i$ 
image zoom of \og\/ position (dashed cross). The reddish pattern is due to strong fringing of the $i\#705$ filter.}
\label{fig:fc}
\end{figure*}

The observational and physical parameter space of known supernova (SN) types has recently been expanded by the discovery of 
unusual optical transients. 
They are fainter and evolve on shorter timescales than normal Type Ia supernovae (SNe Ia) but are brighter than classical novae. Among them are bright and rapidly decaying objects like SN2002bj \citep{poz10}, SN2010X \citep{kal10} and slower evolving ``Ca-rich gap"
transients \citep{ka12} with M$_{R}\gtrsim-16$ mag, of which SN2005E represents the prototype \citep{per10}.
These objects are characterised by strong Ca~{\sc ii} features in their spectra, but their physical origin remains unexplained.
It was initially suggested that some may be associated with the detonation of a helium layer on a low-mass CO white dwarf (WD). Helium outer layers may build up in binary systems in which a primary CO WD accretes from a degenerate (or semi-degenerate) He donor. If a sufficiently massive He layer is accreted ($\gtrsim$0.1~\M), detonation may occur. In the case of a low-mass CO WD, this may lead to a faint thermonuclear SN, roughly one tenth the luminosity of a typical SN Ia luminosity, hence dubbed ``.Ia" \citep{bi07,sh09,shm14}.
A detonation of the He layer may also trigger a secondary detonation of the CO core, in a so-called ``double detonation'' scenario  
 \citep[e.g.][]{fi07,fi10,kr10,woo11,sim12,sh14}.
Both scenarios show titanium and calcium dominated spectra, peak magnitudes of $-19\lesssim{\rm M}_{R}\lesssim-17$  together with rapid lightcurve declines \citep{sh09,sim12}.

However, since none of these transients have unambiguously matched the theoretical expectations of pure helium shell detonation or double detonation, alternative scenarios have also been proposed.
A large explosion of a massive star with several \M\/ of ejected mass may account for SN2010X and SN2002bj \citep{kl14}. The same interpretation is suggested for the rapidly evolving Type Ic SN2005ek \citep{dr13}. Similarly, the helium shell detonation scenario was suggested for SN2005E \citep{per10,wa11}, but alternatives have been proposed for SN2005E and Ca-rich gap transients \citep[see][and reference therein]{kawa10,val14}.

Here, we present the evolution of OGLE-2013-SN-079, a fast evolving SN with peak magnitude that lies in the luminosity window $-19\lesssim{\rm M}_{\rm R}\lesssim-15$. It differs from previous transients in showing titanium-dominated spectra and it appears to be a very promising helium detonation candidate.

\section{Observations}\label{sec:obs}
OGLE-2013-SN-079 (hereafter \og) was discovered by the OGLE-IV Transient Detection System  \citep{wy14}, at m$_{I}\sim21.3$ mag, on 2013 September 30.18 ( magnitudes are reported in Tab.~\ref{table:snm}). The object coordinates have been measured on our astrometrically calibrated images: $\alpha = 00^{\rm h}35^{\rm m}10^{\rm s}.31 \pm0^{\rm s}.05$, $\delta = -67^{\rm o}41'08''.51 \pm0''.05$ (J2000). A spectrum, taken at the New Technology Telescope (NTT) + 
EFOSC2 on October 4.33 {\sc ut},
as part of the Public ESO Spectroscopic Survey for Transient Objects (PESSTO)\footnote{www.pessto.org}, showed a red continuum with broad lines similar to Type I SNe \citep{ch13}. The Galactic reddening toward the SN position is $E(B-V)\sim0.02$ mag \citep{sf11}. Since the available spectra do not show Na~{\sc id} lines related to internal reddening we will assume that the total reddening is given by the Galactic contribution.

\begin{figure*}
\includegraphics[width=18cm]{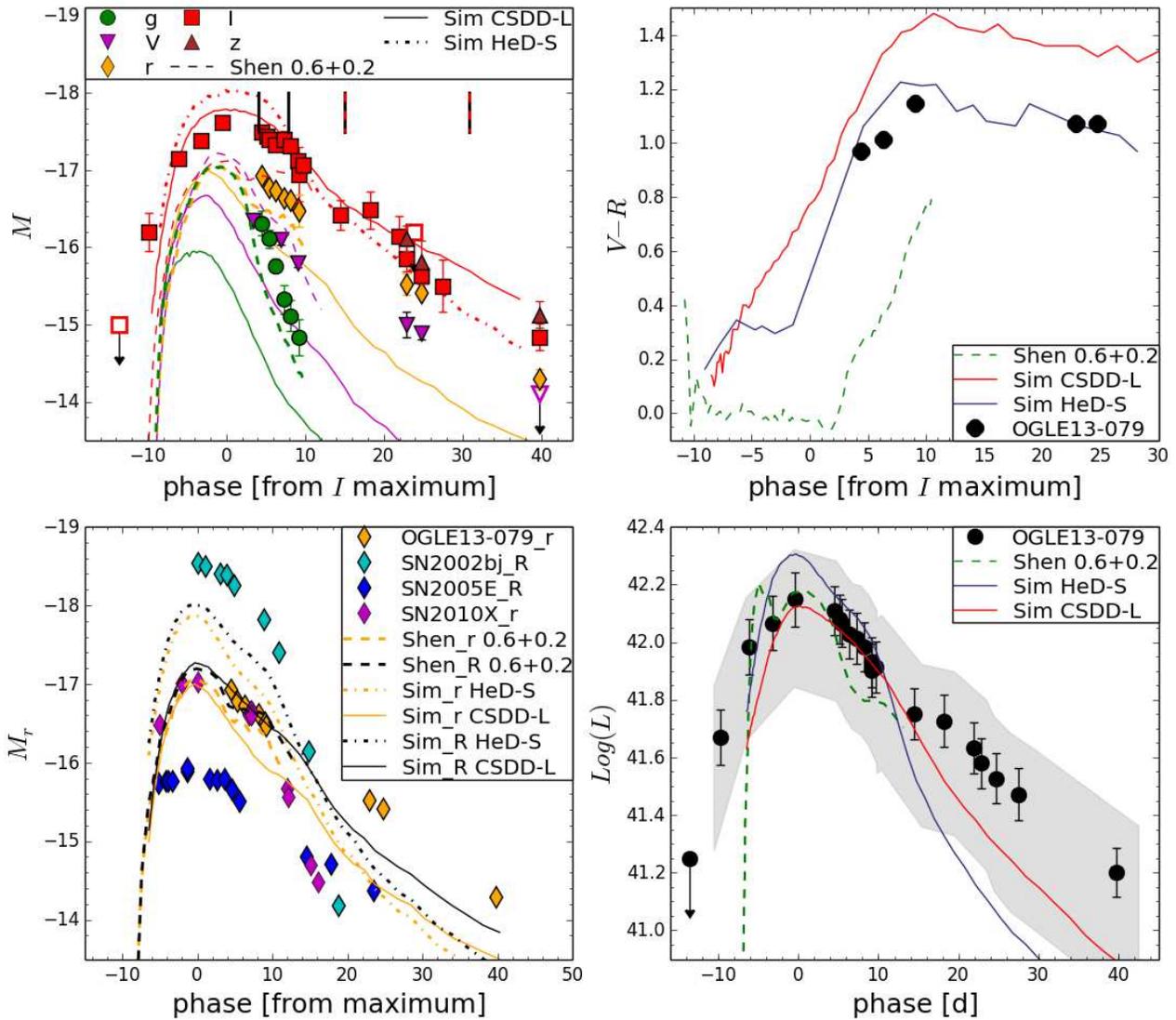}
\caption{Top left: lightcurve evolution in optical bands compared with models. Open symbols denote limits. The vertical dashes indicate epochs with spectroscopic coverage (black optical, red NIR). Top right: $V-R$ evolution compared with models. Bottom left: \og\/ $r$-band evolution compared with those of other transients. Bottom right: Bolometric lightcurve of \og\/ compared with models. The shady grey area indictaes the range of bolometric lightcurves under different assumptions.} 
\label{fig:abs}
\end{figure*}

\subsection{Host and SN location }\label{sec:sn}

On October 16.21 {\sc ut}, we took PESSTO spectra of the two elliptical galaxies, probably interacting, at
coordinates  
$\alpha = 00^{\rm h}35^{\rm m}04^{\rm s}.47$, $\delta = -67^{\rm o}41'02''.7$ (triangle in Fig.~\ref{fig:fc}) and  $\alpha = 00^{\rm h}35^{\rm m}05^{\rm s}.21$, $\delta = -67^{\rm o}41'14''.7$ (square in Fig.~\ref{fig:fc}), finding redshifts $z=0.074$ and $z=0.072$, respectively. The second is also listed in the NASA/IPAC Extragalactic Database as 2MASXJ00350521-6741147 at redshift  $z=0.071$. 
If one of the two galaxies is the host, the transient would be at a projected distance of $\sim49.9$ kpc ($\sim$35") from the first galaxy, or $\sim40.2$ kpc ($\sim$30") from the second.

In our deepest images ($VrI$), we performed aperture photometry on the possible host (the closest).  
We found that the radius containing half of the light of the galaxy is R$_{1/2}\sim28$ kpc, similar to the value  R$_{1/2}\sim30$ kpc measured using the $V$ magnitude of the galaxy and a S\'ersic index of $n=4$ \citep[][and reference therein]{gr13}. Thus \og\/ is likely located in the extreme outskirts of the host, in the halo region. Based on this identification, we adopt $z=0.07$ as the redshift for \og\/ throughout the following.

The host galaxy type and the remote location of \og\/  would argue against a massive star origin of this explosion \citep{ha96,joe09}. The offset distribution of 520 SNe published by \citet{ka12} supports this idea. Massive stars could reach large distances as a consequence of tidal stripping during galaxy interaction but they should then be in the intracluster environment of tidal tails. We do not observe such tails in our data. 

We note that the host and location of \og\/ are different from the bright and fast objects SNe 2002bj and 2010X, which both occurred close to the galaxy nuclei \citep{poz10,kal10}. On the other hand, Ca-rich objects like SN2005E 
were found far from the nucleus \citep{ka12,yu13,ly14}.
\og\/ shows the largest projected distance from its host nucleus among these types of objects and it is comparable only to PTF09dav \citep{su11}.

\subsection{Data}\label{sec:data}
Optical and near infrared (NIR) images were reduced (trimmed, bias subtracted and flat-fielded) using the PESSTO \citep{sm14}, the OGLE \citep{wy14} and Faulkes Telescope pipelines.  Photometric zero-points and color terms were computed using observations of standard fields ($VI$ in Vega and $grz$ in AB system).  
We then calibrated the magnitudes of a local stellar sequence shown in Fig.~\ref{fig:fc}. The average magnitudes of the local-sequence stars were used to calibrate the photometric zero-points in non-photometric nights. The $JHK$ photometry
was calibrated to the Two Micron All Sky Survey system using the same local sequence stars. 
Our optical and NIR photometric measurements were performed using the point-spread function (PSF) fitting technique. 
Differences between passbands were taken into account \citep[applying the S-correction; ][]{str02,pi04}.

All spectra (Table~\ref{table:sp}) were reduced and calibrated in the standard fashion (including trimming, overscan, bias correction, and flat-fielding) using standard routines within {\sc iraf}.  The final flux calibration was checked by comparing the integrated spectral flux, transmitted through standard Sloan or Bessell filters, with our photometry. We applied a multiplicative factor when necessary, and the resulting flux calibration is accurate to within 0.2 mag. 

\section{Photometry}\label{sec:ph}

\subsection{Light curves}
\og\/ was observed during the rising phase only in the $I$-band but subsequently followed in a range of optical filters until it disappeared beyond our detection threshold in late November (Fig.~\ref{fig:abs}). The non-detection four days before the first observation places a strong constraint on the explosion epoch, allowing us to define the rise time and the initial shape of the lightcurve. We estimate the explosion to have occurred at MJD 56553.5$\pm$2, hence $\sim$11 days before I-band maximum. 
The post-peak decline has the same shape in each band with the exception of the $g$-band, in which \og\/ fades by 1.5 mag over five days, in contrast to an average decrease of 0.48 mag in the other bands. The slope in the $VrIz$ bands is fairly constant until the last detection, with the exception of a shoulder in the I-band at $\sim$15 days after peak that is reminiscent of the secondary maximum in type Ia \citep{ka07}. 

The \og\/  magnitudes are comparable to SN2010X but $\sim$1.4 mag
fainter than SN2002bj and $\sim$1.4 mag brighter than SN2005E. 
Indeed, it is brighter than the most luminous \citep[M$_{R}=-16.4$ mag for PTF09dav,][]{ka12} of the
Ca-rich transients. 
However,  the post-peak decline of \og\/ of $\Delta m_{15}(r)\sim$1.2 is similar
to that of SN2005E ($\Delta m_{15}(R)\sim$1.2) and slower than
$\Delta m_{15}(R)\sim$2.4 and $\sim$2.3 of SN2002bj and SN2010X,
respectively.
Thus, while the \og\/ $r$-band peak magnitude is similar to those of the fast evolving objects, the decline is comparable with the slower ones (Fig.~\ref{fig:abs}).

\subsection{Bolometric light curve}
To compare  \og\/ data with theoretical predictions from explosion models (see Section~\ref{sec:mod}),
a bolometric lightcurve of \og\/ was constructed using similar
methods to \citet{in13a,in13b} and is shown in
Fig.~\ref{fig:abs}. 
We initially built a $gVrIz$ pseudo-bolometric lightcurve and added
a $u$ contribution by assuming $u-g\approx0$, which is consistent with the synthetic photometry
retrieved from our early spectra. 
We added a NIR contribution assuming $V-J\approx1.8$ to be constant
throughout the evolution and adopting NIR colors ($J-H$ and $J-K$) similar to the average of SNe
Ia \citep{co10,fr14}. Clearly, the lightcurve constructed via these steps has
substantial uncertainties, therefore, we
also construct a robust lower limit on the bolometric lightcurve by
integrating only over the measured flux in $gVrIz$. We 
placed an
upper limit on the bolometric lightcurve by using the best-fit blackbody curve of the spectral energy distribution retrieved through the red optical bands ($rIz$) by integrating from near ultraviolet to K-band.

The peak luminosity L$_{\rm bol}\approx1.4\times10^{42}$ erg s$^{-1}$ is roughly a factor ten and two less than in normal and faint (SN1991bg-like) SNe Ia, respectively.

\begin{figure*}
\includegraphics[width=18cm]{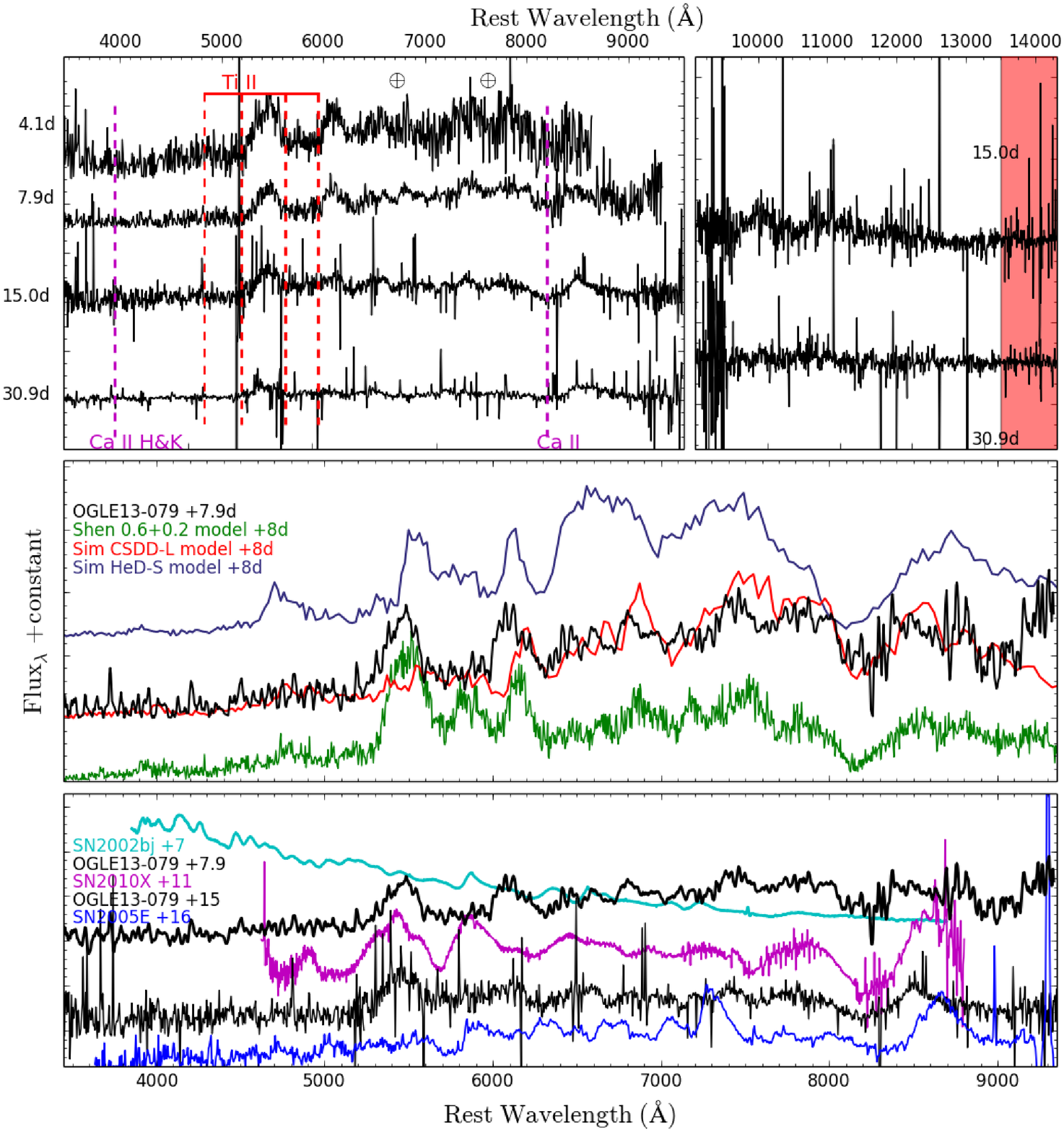}
\caption{Top: Optical (left) and NIR (right) spectral evolution of \og\/. The $\oplus$ symbols mark  
the strongest telluric absorptions. X-SHOOTER spectra are convolved with a Gaussian function of FWHM$=5$\AA\/  and subsequently binned to 5\AA\/ per pixel. 
Middle: comparison between \og\/ spectrum and models at $\sim8$d after $I$-band maximum ($\sim18$d from explosion). Bottom: Comparison between \og\/ spectrum and other transients. } 
\label{fig:spev}
\end{figure*}

\subsection{Models and theoretical light curves}\label{sec:mod}

We compared our photometric and spectroscopic data with those predicted for the detonation of a 0.20\M\/ He layer around a 0.60\M\/ CO core presented by \citet[][hereafter 0.6+0.2\M]{sh10}.
We also compare to angle-averaged predictions from 2D models computed by \citet{sim12}: specifically, we chose one of their helium shell detonations (HeD-S, which has a 0.59\M\/ CO core plus 0.21\M\/ He layer) and one of their double detonations (CSDD-L, which has a 0.59\M\/ CO core plus 0.21\M\/ He layer). These were chosen because they give the best match to \og\/ absolute magnitude. 

Of the models we considered, the best overall fit to the data is found with the CSDD-L model. 
However, the match is not perfect and, with the exception of $I$-band, the model is always too
faint.
We note that a good match of the $I$-band evolution is also achieved by the HeD-S model but this model
has a brighter and wider peak than the data or the CSDD-L model.  
The 0.6+0.2\M\/ model of Shen et al. is fainter and has a faster decline compared to
the HeD model, $\sim$8-10d, depending on the band. This is likely
 a consequence of differences in the model ejecta structure and
 ionization state, which lead to differing degrees of line
 blanketing in the blue and reprocessing of the UV/blue light by heavy-elements.
The 0.6+0.2\M\/ model does produce
an $I$-band shoulder, although $\sim 10$ days earlier than observed. 
We note that the rise time of \og\/  is similar to those
predicted by the models considered, while the ``late'' ($>$15d)
decline has a comparable slope to the Sim et al. the models.
It is also noticeable that SNe 2002bj and 2010X are too rapidly fading compared to the models, while SN2005E is too dim. 

The best match of the bolometric
lightcurve is also found with the CSDD-L model, which is able to adequately
fit the data from -6d to 16d from maximum. After that, \og\/ declines
more slowly for the next 10 days ($15\lesssim {\rm d}\lesssim25$) but
then settles onto a decline slope similar to the model.
The HeD-S model is between the limits of the
bolometric lightcurves (grey area in Fig.~\ref{fig:abs})
, but
the fit generally appears poorer both around peak and at later times
compared to the CSDD-L model. The Shen et al. model also fits the data
reasonably well, although in the bolometric lightcurve of \og\/  we do not observe any double-peaked behavior. 
Rise time and post maximum evolution differences between data and models are roughly similar to the single bands previously shown.

\subsection{Color evolution comparison}
In the top right panel of Fig.~\ref{fig:abs} we compare the
$V-R$ color evolution of \og\/ with those predicted by the models. Since we
did not have $R$-band observations, we transformed $r$ to $R$ magnitude
using {\sc snake} (a {\sc python} code for $K$-correction and
magnitude conversion, Inserra in preparation). Although
the lightcurves are more similar to the CSDD-L model (see above), the
$V-R$ behavior is closer to the HeD-S model with a $V-R\sim$1.1 after 10 days since maximum. We note that the $V-R$
evolution is also comparable to 
CSDD-L model but
shifted by  
$\sim$0.3 to bluer values.   Both the Sim et al. and Shen et al. models of pure
helium detonation have similar behavior, although the Sim et al.  models are systematically redder\footnote{The redder colors of the
Sim et al. compared to Shen et al. models are likely attributable to a combination of differences in the treatment of ionization and ejecta composition.}.

\section{Spectroscopy}\label{sec:sp}

The rest-frame spectra are shown in Fig.~\ref{fig:spev} 
and reported in Tab.~\ref{table:sp}. 
The striking features of the spectral sequence are the red color
and the two prominent absorption profiles at 5300\AA\/ and
5900\AA\/. To our knowledge such features have never been observed at an
early phase in any other SN. By comparison with wavelengths and oscillator strengths from the Kurucz database \citep{kur} we identified these as multiple, blended Ti lines. 
 Similar lines have been predicted in helium shell detonation models
 \citep{sh10,wa11}. The spectra are dominated by Ti~{\sc ii} lines
 during the 30d post peak evolution; the majority of the line
 profiles bluer than 6000\AA\/ are related to Ti~{\sc ii} lines\footnote{ We note that titanium production could be important for the positron budget and 511 kev emission \citep{per14}}, with
 the exception of the weak, or possibly absent, Ca H\&K lines. 
We note that sub-luminous type Ia also show strong Ti absorptions in the blue. 
The Ca~{\sc ii} NIR triplet becomes more prominent from the second
spectrum onwards. 
However, it is weaker than the 5300\AA\/ line and weaker than typically observed in Ca-rich objects. In these objects the Ca~{\sc ii} is roughly three times stronger than other features \citep[see][]{val14}. 
The NIR absorption around 9800\AA\/ could be attributed either to Ti~{\sc ii} line 
or Ca~{\sc ii}. 
We do not observe any obvious lines associated with intermediate elements
such as Si, S and light elements as Mg, C and O as seen in other
possible helium shell detonation candidates.
Ti~{\sc ii} and Ca~{\sc ii} are dominant, and consistent with what is expected in a helium shell detonation \citep[see also][]{hol13}.
We note that \og\/ is the first transient showing strong Ti lines, hence it could be the first example of what might be called  a ``Ti-strong" object.

\subsection{Comparison with other helium shell detonation candidates}\label{ss:spcm}
In 
Fig.~\ref{fig:spev} we show
\og\/ with three other well studied 
 transients of similar luminosity, 
SNe 2002bj, 2005E and 2010X, at similar epochs. 
The spectra of SN2002bj are quite different from
those of \og\/: SN2002bj has a much bluer
continuum and does not show the strong Ti lines (around 5300\AA\/ and
5900\AA). 
SN2005E has a similar red color to \og\, but also does not show the
two prominent Ti lines. 
In contrast, SN2005E shows a strong Ca\,{\sc ii} NIR triplet feature, forbidden [Ca\,{\sc ii}] 7291,7323\AA , 
and unambiguous He\,{\sc i} lines. These led \citet{per10}  to identify the object as a 
Type Ib and a Ca-rich transient. We do not see forbidden [Ca\,{\sc ii}] or any He\,{\sc i} lines 
in \og\/ spectra at comparable phases. \og\/ has a similar colour to SN2010X but 
the spectral features are again different. The spectra are particularly distinct between 5500\AA\
and 7500\AA\ where Na\,{\sc i},  O\,{\sc i} and Mg\,{\sc i} are not present in \og.
SN2010X also shows a prominent Ca\,{\sc ii} NIR triplet, which is similar in strength to those of other Ca-rich events
but is not visible in the \og\/ spectrum with comparable intensity. 
We note that the \og\/ Ca\,{\sc ii} NIR triplet velocity is $\sim4000$~km/s slower than those of SNe 2005E and 2010X in the same phase.
We conclude that \og\/ does not closely resemble any of these objects
in its spectral features or colors. 

\subsection{Comparison with double-detonation and helium shell detonation models}

In 
Fig.~\ref{fig:spev} we compare the \og\/ spectrum with Sim et al. and Shen et al.
models at similar epochs ($\sim18$ days from explosion). 
The flux computed from the model spectra has been scaled to match that of \og.
All lines and their absorption and emission strength are well
reproduced by the Shen et al. 0.6+0.2\M\/ model. The model is
slightly bluer, as also highlighted by the color evolution. The CSDD-L model does a reasonably good job in reproducing the spectrum,
although we note that it does not provide a good match to the two
noteworthy lines at 5300\AA\/ and
5900\AA\/, which we have attributed to Ti~{\sc ii}.
These differences between the models could be due to the distribution of Fe-group elements in the ejecta, which 
are key to shaping the spectra \citep[see Fig. 6 of ][]{sim12},
 and generally to the treatment of ionization.
The comparison between the 0.6+0.2 and HeD-S models (similar masses but different densities) shows that the Sim et al. models fit the color better but the Shen et al. model fit the Ti~{\sc ii} lines quite well. 
The existing models do not predict strong He lines, however we note that further studies that include treatments of non-thermal excitation are required to better investigate the formation of He lines in the optical \citep{hac12}.
We note that the majority of model lines are either not observed or have different strength (e.g. Ca~{\sc ii}) in the other transients mentioned in Section~\ref{ss:spcm}. 
Although \og\/ lightcurves could, in principle, be fitted with a massive star explosion it would be difficult to reproduce an oxygen-free and titanium-dominated spectral sequence.

\section{Conclusions}\label{sec:conc}

\og\/ is a transient in the luminosity region $-19\lesssim{\rm M}_{\rm R}\lesssim-15$ spectrophotometrically different from similar transients discovered so far.
Its maximum luminosity M$_{\rm I}\sim-17.6$  is similar to SN2010X, thus
brighter than the prototype Ca-rich transient SN2005E  by $\sim$1.5 mag and less luminous than
SN2002bj by $\sim$1 mag. 
Its
decline is comparable to that of SN2005E and other Ca-rich transients. 
Its location in the outskirts of the host galaxy ($\sim$40 kpc), similar only to the Ca-rich object PTF09dav, makes \og\/ the most remote gap object so far discovered. 
\og's spectral evolution is unique since there is little or no trace of C, O, Mg, Si, S
in contrast to 
the other transients 
like SNe 2002bj, 2005E and 2010X. The \og\/
spectra are dominated by He-burning products such as Ti~{\sc ii} and
Ca~{\sc ii}. The titanium lines are noticeably 
stronger than any other elements and suggests that \og\/  could be the first ``Ti-strong" transient.

\og\/ is the first transient that reasonably well matches the synthetic observables predicted by models for detonation of a He layer around a low-mass CO WD and/or equivalent double detonation models. 
The strong lines of Ti\,{\sc ii}, a He-burning product, are reproduced by the nucleosynthetic
reactions in these models. The lack of prominent features from intermediate-mass elements
in the spectra of OGLE13-079 is striking. This combination is a quantitative signature
of these explosion models.  
Not all the observed characteristics of OGLE13-079 are well matched by these models. 
More focused and detailed theoretical simulations are warranted to investigate if the full data 
set can be matched with a pure helium shell detonation or a double detonation of a lower-mass shell ($<0.2$ \M).

\acknowledgments

CI thanks 
M.Kasliwal (SN2010X data).
Based on observations at ESO as part of PESSTO (188.D-3003/191.D-0935/092.D-0555).
Funded by FP7/2007-2013/ERC Grant agreement [291222](SJS).
We acknowledge support from STFC grant ST/L000709/1(SJS,SS), TRR33 grant of DFG(ST),
FP7/2007-2013 grant [267251](NER), FP7/ERC grant [320360](MF).

\begin{table*}
\caption{Observed photometry of \og\/ and assigned errors.}
\begin{center}
\begin{tabular}{ccccccccc}
\hline
\hline
Date & MJD & Phase* & g & V & r & I & z & Telescope\\
dd/mm/yy &  & (days)& & & & & &\\
\hline
16/09/13  &   56551.15  &-14.0&  &      &  &   $>$22.48 &  &  OGLE   \\
20/09/13 &     56555.18 &-10.0&     &    &   &   21.28 (0.25)&  &   OGLE\\ 
24/09/13  &   56559.15  &-6.0&    &    &   &   20.34 (0.07)&  &   OGLE\\ 
27/09/13  &    56562.16  &-3.0&    &    &   &   20.10 (0.05)&  &   OGLE\\ 
30/09/13  &    56565.19 &0.0&      &    &  &   19.87 (0.04)&  &  OGLE\\ 
03/10/13 &  56569.32  &4.1&    & 21.16 (0.06)  &       &   &  & NTT\\
05/10/13&  56570.43&5.2&  	    21.21 (0.16)&&  20.58 (0.03)&  19.99 (0.04) & & FTS\\
06/10/13 &   56571.16  & 6.0&     &  &   &  20.05 (0.06)&  &  OGLE\\ 
06/10/13 & 56571.42&6.2 	&    21.41 (0.12)& & 20.72 (0.06)&  20.09 (0.04) &  & FTS\\
07/10/13 & 56572.42& 7.2& 	    21.76 (0.08)& & 20.77 (0.02) &  20.16 (0.04)  &  & FTS\\
07/10/13&  56573.03  &7.8& &  21.40 (0.06)     &   &   &   & NTT\\
08/10/13 & 56573.50&8.3&  	    22.20 (0.18)& & 20.86 (0.07) & 20.09 (0.10)& 	&  FTS\\
09/10/13 & 56574.44&9.2&  	    22.41 (0.20)& & 20.89 (0.04)&  20.18 (0.08)&  & FTS\\
09/10/13 & 56575.38&10.2&&  	21.71 (0.05)   &  20.99 (0.06) & 20.37 (0.06) & & NTT\\
10/10/13 & 56575.50&10.3&      22.69 (0.23)&&  21.03 (0.20)&  20.54 (0.15)&  &  FTS\\
11/10/13   &   56576.10 &10.9&       &     &   &   20.45 (0.08)&  &    OGLE\\ 
16/10/13  &    56581.13 &15.9&   &   &   &   21.06 (0.20)&  &    OGLE\\ 
20/10/13  &    56585.18 &20.0&    & &   &    21.00 (0.24)&&    OGLE\\ 
24/10/13  &    56589.20 &24.0&    & &   &    21.34 (0.26)&  &    OGLE\\ 
24/10/13&  56590.17 &25.0&    & 22.51 (0.16) &   21.98 (0.14)&  21.63 (0.16)&  21.36 (0.16)&  NTT\\
26/10/13 &     56591.07 & 25.9&    &   &   &   $>$ 21.28 & &    OGLE \\
26/10/13&  56592.19  &27.0&    & 22.62 (0.08)  & 22.09 (0.05) & 21.86 (0.08)&  21.67 (0.18)&  NTT\\
30/10/13  &    56595.11  &29.9&    &    &   &  21.99 (0.34)&    &  OGLE\\ 
07/11/13 &    56603.08& 37.9 &   &     &  &  $>$22.00&   &  OGLE \\
11/11/13 & 56608.25  & 43.1 & $>$23.40 &  &  23.20 (0.13)&  22.65 (0.17) & 22.35 (0.17) & NTT\\
Deepest limits &   & &  &  &  &   &  & \\
25/11/13 & 56622.15  & 57.0 &  & $>$23.45 &  &   & & NTT\\
01/12/13 & 56628.11  & 62.9 & $>$23.50 & & $>$24.00 &  $>$24.20 &$>$23.50 & NTT\\
\hline
\hline
Date & MJD & Phase* & & $J$ & $H$ & $K$ & & Telescope  \\
dd/mm/yy &  & (days)  &  & &\\
\hline
13/10/13 &  56579.21  &14.0 && 	 19.98  (.20) &    $>$19.60 & 	  $>$ 19.20& & NTT\\
25/10/13&   56591.37 &  26.2&& 	 $>$20.10      &  	&   &   & NTT\\  
02/11/13 &  56599.31  & 34.1 & &  $>$20.00  &   &  &    & NTT\\
\hline
\end{tabular}
\end{center}
*Phase with respect to the $I$-band maximum. 
\label{table:snm}
\end{table*}%

\begin{table*}
\caption{Spectroscopic observations.}
\begin{center}
\begin{tabular}{cccccl}
\hline
\hline
Date & MJD & Phase* &Range & Resolution & Instrumental  \\
dd/mm/yy &  & (days)  &(\AA)  & (\AA)  &Configuration\\
\hline
03/10/13 & 56569.32 & 4.1 &3700--9300 & 18 & NTT+EFOSC2+gm13\\                  
07/10/13 & 56573.05       & 7.9   & 3400--10300  & 11/16 & NTT+EFOSC2+gm11/gm16\\                                                                           
15/10/13 & 56580.18       &  15.0  &  3400--23200  & 0.6/2 & VLT+XSHOOTER+VIS/NIR    \\                                                                                
30/10/13 & 56596.08        &  30.9  & 3400--23200   &  0.6/2& VLT+XSHOOTER+VIS/NIR \\                                                                                                                            
\hline
\end{tabular}
\end{center}
*Phase with respect to the $I$-band maximum. 
\label{table:sp}
\end{table*}%


\begin{thebibliography}{}
\label{refs}
\bibitem[Anderson 
\& James(2009)]{joe09} Anderson, J.~P., \& James, P.~A.\ 2009, \mnras, 399, 559
\bibitem[Bildsten et al.(2007)]{bi07} Bildsten, L., Shen, 
K.~J., Weinberg, N.~N., \& Nelemans, G.\ 2007, \apjl, 662, L95
\bibitem[Chen et al.(2013)]{ch13} Chen, T.-W., Inserra, C., 
Nicholl, M., et al.\ 2013, The Astronomer's Telegram, 5443, 1
\bibitem[Contreras et al.(2010)]{co10} Contreras, C., Hamuy, 
M., Phillips, M.~M., et al.\ 2010, \aj, 139, 519
\bibitem[Drout et al.(2013)]{dr13} Drout, M.~R., Soderberg, 
A.~M., Mazzali, P.~A., et al.\ 2013, \apj, 774, 58
\bibitem[Fink et 
al.(2007)]{fi07} Fink, M., Hillebrandt, W., R\"opke, F.~K.\ 2007, \aap, 476, 1133 
\bibitem[Fink et 
al.(2010)]{fi10} Fink, M., R{\"o}pke, F.~K., Hillebrandt, W., et al.\ 2010, \aap, 514, A53 
\bibitem[Friedman et al.(2014)]{fr14} Friedman, A.~S., 
Wood-Vasey, W.~M., Marion, G.~H., et al.\ 2014, arXiv:1408.0465
\bibitem[Graham(2013)]{gr13} Graham, A.~W.\ 2013, Planets, 
Stars and Stellar Systems.~Volume 6: Extragalactic Astronomy and Cosmology, 
91 
\bibitem[Hachinger et al.(2012)]{hac12} Hachinger, S., 
Mazzali, P.~A., Taubenberger, S., et al.\ 2012, \mnras, 422, 70
\bibitem[Hamuy et al.(1996)]{ha96} Hamuy, M., Phillips, 
M.~M., Suntzeff, N.~B., et al.\ 1996, \aj, 112, 2391 
\bibitem[Holcomb et al.(2013)]{hol13} Holcomb, C., 
Guillochon, J., De Colle, F., \& Ramirez-Ruiz, E.\ 2013, \apj, 771, 14 
\bibitem[Inserra et 
al.(2013a)]{in13a} Inserra, C., Pastorello, A., Turatto, M., et al.\ 2013, \aap, 555, A142 
\bibitem[Inserra et al.(2013b)]{in13b} Inserra, C., Smartt, 
S.~J., Jerkstrand, A., et al.\ 2013, \apj, 770, 128
\bibitem[Kasen 
\& Plewa(2007)]{ka07} Kasen, D., \& Plewa, T.\ 2007, \apj, 662, 459 
\bibitem[Kasliwal et al.(2012)]{ka12} Kasliwal, M.~M., 
Kulkarni, S.~R., Gal-Yam, A., et al.\ 2012, \apj, 755, 161
\bibitem[Kasliwal et al.(2010)]{kal10} Kasliwal, M.~M., 
Kulkarni, S.~R., Gal-Yam, A., et al.\ 2010, \apjl, 723, L98
\bibitem[Kawabata et al.(2010)]{kawa10} Kawabata, K.~S., 
Maeda, K., Nomoto, K., et al.\ 2010, \nat, 465, 326 
\bibitem[Kleiser 
\& Kasen(2014)]{kl14} Kleiser, I.~K.~W., \& Kasen, D.\ 2014, \mnras, 438, 318
\bibitem[Kromer et al.(2010)]{kr10} Kromer, M., Sim, S.~A., 
Fink, M., et al.\ 2010, \apj, 719, 1067
\bibitem[Kurucz 
\& Bell(1995)]{kur} Kurucz, R., \& Bell, B.\ 1995, Atomic Line Data (R.L.~Kurucz and B.~Bell) Kurucz CD-ROM No.~23.~Cambridge, Mass.: Smithsonian Astrophysical Observatory, 1995., 23,  
\bibitem[Lyman et al.(2014)]{ly14} Lyman, J.~D., Levan, 
A.~J., Church, R.~P., Davies, M.~B., 
\& Tanvir, N.~R.\ 2014, \mnras, 444, 2157 
\bibitem[Perets(2014)]{per14} Perets, H.~B.\ 2014, 
arXiv:1407.2254
\bibitem[Perets et al.(2010)]{per10} Perets, H.~B., Gal-Yam, 
A., Mazzali, P.~A., et al.\ 2010, \nat, 465, 322
\bibitem[Pignata et al.(2004)]{pi04} Pignata, G., Patat, F., 
Benetti, S., et al.\ 2004, \mnras, 355, 178
\bibitem[Poznanski et al.(2010)]{poz10} Poznanski, D., 
Chornock, R., Nugent, P.~E., et al.\ 2010, Science, 327, 58
\bibitem[Schlafly 
\& Finkbeiner(2011)]{sf11} Schlafly, E.~F., \& Finkbeiner, D.~P.\ 2011, \apj, 737, 103
\bibitem[Shen 
\& Moore(2014)]{shm14} Shen, K.~J., \& Moore, K.\ 2014, \apj, 797, 46
\bibitem[Shen 
\& Bildsten(2014)]{sh14} Shen, K.~J., \& Bildsten, L.\ 2014, \apj, 785, 61
\bibitem[Shen et al.(2010)]{sh10} Shen, K.~J., Kasen, D., 
Weinberg, N.~N., Bildsten, L., \& Scannapieco, E.\ 2010, \apj, 715, 767 
\bibitem[Shen 
\& Bildsten(2009)]{sh09} Shen, K.~J., \& Bildsten, L.\ 2009, \apj, 699, 1365
\bibitem[Sim et al.(2012)]{sim12} Sim, S.~A., Fink, M., 
Kromer, M., et al.\ 2012, \mnras, 420, 3003
\bibitem[Smartt et al.(2014)]{sm14} Smartt, S.~J., Valenti, 
S., Fraser, M., et al.\ 2014, arXiv:1411.0299 
\bibitem[Stritzinger et al.(2002)]{str02} Stritzinger, M., 
Hamuy, M., Suntzeff, N.~B., et al.\ 2002, \aj, 124, 2100 
\bibitem[Sullivan et al.(2011)]{su11} Sullivan, M., 
Kasliwal, M.~M., Nugent, P.~E., et al.\ 2011, \apj, 732, 118 
\bibitem[Valenti et al.(2014)]{val14} Valenti, S., Yuan, F., 
Taubenberger, S., et al.\ 2014, \mnras, 437, 1519
\bibitem[Waldman et al.(2011)]{wa11} Waldman, R., Sauer, D., 
Livne, E., et al.\ 2011, \apj, 738, 21
\bibitem[Woosley 
\& Kasen(2011)]{woo11} Woosley, S.~E., \& Kasen, D.\ 2011, \apj, 734, 38
\bibitem[Wyrzykowski et al.(2014)]{wy14} Wyrzykowski, L., 
Kostrzewa-Rutkowska, Z., Kozlowski, S., et al.\ 2014, AcA, 64, 197
\bibitem[Yuan et al.(2013)]{yu13} Yuan, F., Kobayashi, C., 
Schmidt, B.~P., et al.\ 2013, \mnras, 432, 1680

\end{thebibliography}
\end{document}